\documentclass[conference]{IEEEtran}
\IEEEoverridecommandlockouts
\usepackage{cite}
\usepackage{subfigure}
\usepackage{amsmath,amssymb,amsfonts}
\usepackage{algorithmic}
\usepackage{graphicx}
\usepackage{textcomp}
\usepackage{balance}
\usepackage{xcolor}
\usepackage{hyperref}
\usepackage{tikz}
\usepackage{booktabs}
\tikzstyle{mybox} = [draw=black, very thick, rectangle, rounded corners, inner ysep=5pt, inner xsep=5pt]

\usepackage{multirow} 
\newcommand{\celllinebreak}[2][t]{%
  \begin{tabular}[#1]{@{}l@{}}#2\end{tabular}}

\def\BibTeX{{\rm B\kern-.05em{\sc i\kern-.025em b}\kern-.08em
    T\kern-.1667em\lower.7ex\hbox{E}\kern-.125emX}}

\begin{document}

\title{Run, Forest, Run? On Randomization and Reproducibility in Predictive Software Engineering}

\author{\IEEEauthorblockN{Cynthia C.~S. Liem}
\IEEEauthorblockA{\textit{Multimedia Computing Group} \\
\textit{Delft University of Technology}\\
Delft, The Netherlands \\
c.c.s.liem@tudelft.nl}
\and
\IEEEauthorblockN{Annibale Panichella}
\IEEEauthorblockA{\textit{Software Engineering Research Group} \\
\textit{Delft University of Technology}\\
Delft, The Netherlands \\
a.panichella@tudelft.nl}
}

\maketitle

\begin{abstract}
Machine learning (ML) has been widely used in the literature to automate software engineering tasks. However, ML outcomes may be sensitive to randomization in data sampling mechanisms and learning procedures.
To understand whether and how researchers in SE address these threats, we surveyed 45 recent papers related to three predictive tasks: defect prediction (DP), predictive mutation testing (PMT), and code smell detection (CSD). We found that less than 50\% of the surveyed papers address the threats related to randomized data sampling (via multiple repetitions); only 8\% of the papers address the random nature of ML; and parameter values are rarely reported (only 18\% of the papers).  
To assess the severity of these threats, we conducted an empirical study using 26 real-world datasets commonly considered for the three predictive tasks of interest, considering eight common supervised ML classifiers. We show that different data resamplings for 10-fold cross-validation lead to extreme variability in observed performance results. Furthermore, randomized ML methods also show non-negligible variability for different choices of random seeds. More worryingly, performance and variability are inconsistent for different implementations of the conceptually same ML method in different libraries, as also shown through multi-dataset pairwise comparison. To cope with these critical threats, we provide practical guidelines on how to validate, assess, and report the results of predictive methods.
\end{abstract}



\section{Introduction}


The application of machine learning (ML) techniques to software engineering (SE) has attracted increasing interest from researchers and practitioners in recent years. This has resulted in various successful ML applications, such as building predictive models for code quality assessment~\cite{Hosseini:survey}, software testing~\cite{zhang2016predictive,zhang2018predictive}, effort estimation, and software security. At the same time, despite increasing optimism about the effectiveness of ML techniques for SE, researchers have started questioning the reproducibility of their results~\cite{mahmood2018reproducibility, Shepperd2014bias}. For example, within the SE community, attention has been raised for best practices to deal with outcomes of randomized algorithms, and to better substantiate whether observed performance differences are true differences through the application of statistical tests~\cite{arcuri2014hitchhiker}. When applying these, it also is important to be explicitly aware of underlying assumptions made by the test~\cite{arcuri2014hitchhiker, herbold2017comments}.


Outside of the SE domain, other fields have demonstrated how sensitive analyses and drawn conclusions may be to \emph{researcher degrees of freedom} that are not commonly reported. In Psychology, warnings have been raised that undisclosed (and non-malicious) choices in data collection, analysis, and reporting can easily lead to false-positive results and conclusions~\cite{simmons2011false}. In Music Information Retrieval research, it was shown that implementation design choices for evaluation metrics can yield highly different outcomes~\cite{raffel2014mireval}, and that common descriptions and commonly reported parameters of music signal processing procedures are insufficiently specific to warrant unambiguous, stable reimplementations~\cite{mcfee_open_2019}.

Despite great and broad optimism about ML's promises, criticism has risen within the ML field itself on the true merit, reasons behind, and reproducibility of reported advances. The tendency to appreciate more complex approaches that outperform alternatives, together with short, fast-paced publication cycles, has been critiqued to lead to flawed works that lack practical use and empirical rigor~\cite{Lipton:2019:RPT:3336127.3316774, sculley2018winnerscurse}. It has been hinted that the broader artificial intelligence faces a reproducibility crisis~\cite{Hutson725}, which may e.g.\ be due to unreleased source code and unreported critical configuration details. In deep reinforcement learning, experiments varying random seeds and hyperparameter values caused major performance differences, and alternative implementations of the same baseline task were shown to give radically different results, leading to calls for more thorough reporting practice~\cite{deeprl18}.


While many ML applications in SE use older, more established supervised ML techniques than those under particular scrutiny today, it still is imaginable that similar issues may be at play, impacting the trustability of observed outcomes. This paper therefore seeks to gain more insight into how rigorously ML techniques are assessed in SE domains. More precisely, we investigate three important factors that can potentially impact ML result variability, threatening future reproducibility: (1) the sensitivity of ML methods to randomized resamplings from datasets for training and validation; (2) the randomized nature of specific ML learning procedures; (3) whether different implementations of the conceptually same ML method in different libraries are actually consistent with one another.

Focusing on three actively researched predictive SE tasks (defect prediction, predictive mutation testing, and code smell detection), we first perform a systematic literature review, to understand the extent to which researchers addressed the factors of our interest in the current literature. Alarmingly, after rigorously analysing 45 recent papers in the three selected SE tasks, we did not find a single case in which all three potential treats have been adequately and explicitly addressed.

Secondly, we present an empirical study that aims to investigate to what extent predictive SE outcomes may be affected by the factors of our interest.
Our study involves 26 commonly-used datasets in SE from the three predictive SE tasks of interest, eight ML classifier models, and three alternative commonly-used ML libraries. Our results show striking inconsistencies and outcome variabilities, that need to be more carefully and explicitly addressed by researchers than seemingly is done today. Therefore, we conclude our work by
 presenting a set of practical guidelines on how to address these threats adequately, and how to validate, assess, and report the results in predictive software engineering.

\section{Systematic Review}
Several current ML criticisms on threats to validity and reproducibility may also hold for common ML practices in SE. First of all, for training and evaluation, samples or data partitions are randomly drawn from a dataset, which may lead to particularly (un)lucky samplings. Next to this, models may include randomization as part of their training procedure (e.g.\ in a random forest), implying that different random seeds may give different results. Finally, hyperparameters have a major impact on outcomes, and thus it would be good practice to report and discuss them. To assess to what extent these aspects are acknowledged in current SE practice, we consider three representative predictive SE tasks that currently heavily build on ML approaches: \emph{defect prediction (DP)}, \emph{predictive mutation testing (PMT)}, and \emph{code smell detection (CSD)}.

\subsubsection{DP} In this task, ML methods are used to predict the fault-proneness of software modules (e.g., class or methods) in a given software project with the goal of prioritizing QA activities. ML for DP has been extensively studied in the last two decades, as reflected in
existing surveys (e.g., \cite{Hosseini:survey}), reflection papers (e.g., \cite{Kamei2016, Wan2018}), and dedicated tools
(e.g., ~\cite{herbold2015crosspare}). 
Defect prediction can be formulated as a binary classification problem, whose goal is to predict whether each specific software module (e.g., a class or method) is defective or not~\cite{Hosseini:survey, Kamei2016, Wan2018}. Multiple features have been used for building prediction models~\cite{Kamei2016}, such as code (e.g.,~\cite{menzies2006data}), process (e.g,~\cite{nagappan2005use}), and organizational metrics (e.g,~\cite{cataldo2009software}).

\subsubsection{PMT} Mutation testing aims to assess the effectiveness of a test suite. Given a program $P$, mutation testing tools generate program variants (i.e., \textit{mutants}) based on a given set of transformation rules (i.e., \textit{mutation operators}). An effective test case should pass when executed against the original program $P$ but fails against a mutant $P'$. In other words, the test should differentiate between the original program and its mutated variant. In this scenario, the mutant $P'$ is said to be \textit{killed}. Instead, the mutant $P
'$ \textit{survives} if the test cases pass on both $P$ and $P'$. The test suite's effectiveness is then measured as the ratio between the number of killed mutants and the total number of non-equivalent mutants being generated~\cite{Jia2011survey}.  While mutation testing is widely recognized as a \textit{high-end} test adequacy criterion~\cite{ammann2016introduction}, it does incur a considerable overhead. Indeed, performing mutation testing requires to rerun the entire test suite against each generated mutant. For this reason, researchers have proposed various strategies that aim to reduce the cost of mutation testing~\cite{offutt2001mutation}.

The purpose of PMT is to \textit{produce the mutation testing results without executing the actual mutants}~\cite{zhang2016predictive,zhang2018predictive}. PMT has been formulated as a binary classification problem, whose goal is to predict whether each mutant $P'$---generated by a mutation testing tool---is killed or not by a given test suite $T$~\cite{zhang2016predictive,zhang2018predictive}. Widely applied features for PMT include test execution data (e.g., the number executions of a mutated statement), infection features (e.g., the type of mutation operator), and propagation features (e.g., the number of assertions that cover a mutant). 

\subsubsection{CSD} Code smells are symptoms of poor software design choices that negatively impact code comprehensibility and future maintenance activities~\cite{sharma2018survey}. The original catalog has also been lately extended to test code~\cite{van2001refactoring}, and databases~\cite{sharma2016does}. Over the last decades, researchers have proposed various static analysis methods and tools to detect code smells automatically~\cite{fernandes2016review}. A complete overview of code smell detection approaches is available in~\cite{sharma2018survey}. 

More recent approaches to detect code smell rely on ML~\cite{arcellifontana2016smell, arcellifontana2017smellseverity,dinucci2018smell}. The detection problem is reformulated as a binary classification task using supervised ML methods. More precisely, the goal is to predict/detect whether a software module (class or method) is affected by a given code smell. These approaches typically use various static software metrics
~\cite{sharma2018survey}, such as lines of code, McCabe cyclomatic complexity, the number of input parameters of a method, and so on.

While not aiming to provide an exhaustive literature review for these three tasks, we still would like a representative snapshot of current common practices, obtained in a rigorous and unbiased fashion. Therefore, we conduct a systematic literature review in line with common practice in the SE domain~\cite{kitchenham2004, kitchenham2009systematic}, as described below.



\subsection{Methodology}
We used digital libraries and search engines (Google Scholar, Scopus, and DBLP) to manually retrieve papers related to the predictive SE tasks of interest. For querying, we considered the following search keywords: 
\begin{itemize}
    \item DP: (``defect prediction'') OR (``fault prediction'') OR (``bug prediction'') OR (``defect'' AND ``deep learning'') OR (``fault'' AND ``deep learning'') OR (``bug'' AND ``deep learning'') OR (``supervised defect prediction'').
    \item PMT: (``predictive mutation testing'') OR (``supervised mutation testing'') OR (``predictive mutation analysis'') OR (``machine learning'' AND ``mutation testing'') OR (``deep learning'' AND ``mutation testing'').
    \item CSD: (``predictive code smell'') OR (``code smell'' AND ``machine learning'') OR (``code smell'' AND ``deep learning'') OR (``design flaw'' AND ``machine learning'') OR  (``design flaw'' AND ``deep learning'') OR (``supervised code smells'').
\end{itemize}
With these keywords initially yielding thousands of hits, we used \textit{computer science} and \textit{software engineering} as additional filters to explicitly restrict to papers relevant to the SE domain. We then restricted the selection further to works that appeared at international SE top venues: IEEE Transactions on Software Engineering (TSE), IEEE Transactions on Reliability, Empirical Software Engineering (EMSE), Information and Software Engineering (IST), Journal and Software and System (JSS), Software Testing, Verification and Reliability (STVR),  IEEE Access, Software Quality Journal (SQJ), ACM/IEEE International Conference on Software Engineering (ICSE), Mining Software Repositories (MSR), IEEE International Conference on Software Testing, Verification and Validation (ICST), Automated Software Engineering (ASE), Symposium on the Foundations of Software Engineering (FSE), ACM International Symposium on Software Testing and Analysis (ISSTA), IEEE International Conference on Software Analysis, Evolution and Re-engineering  (SANER), and International Symposium on Software Reliability Engineering (ISSRE). To focus on original work presentations at sufficient levels of detail, we \textbf{excluded} keynotes, extended abstracts, tutorials, and posters.
The inclusion and exclusion criteria are detailed in the following paragraphs and for each predictive tasks separately. 

\subsubsection{DP}
given the large body of DP research and our focus on a snapshot of current common practices, we focus our search on papers published in the selected top venues in 2019 and 2020. We \textbf{included} papers that discuss the application of ML methods or improved ML procedures (e.g.\ data balancing), framing DP as a binary classification task. Hence, we \textbf{excluded} papers that focus on fault localization, or predicting the severity (or the number of) defects in software modules.

\subsubsection{PMT} given that the first paper was published in 2016~\cite{zhang2016predictive}, the body of research in PMT is still small. For that reason, we included all papers published in the selected venues between the years 2016 and 2020. We \textbf{included} all papers that applied ML approaches to mutation testing and its surrounding challenges (e.g.\ data balancing, transfer learning), while explicitly focusing on the framing of PMT as a binary classification task at the mutant level (as this makes results comparable with those of classical mutation testing). We \textbf{excluded} papers that use ML-based techniques that do not focus on predictions at this level.

\subsubsection{CSD} with ML-based CSD also being a very recent research topic, we consider all papers published in the selected top venues. Beyond this, we include any further works mentioned in a recent systematic literature review~\cite{azeem19smellreview}, and apply \textit{snowballing} as defined by Wohlin~\cite{wohlin2014guidelines}, i.e., including papers cited by or that cite the papers retrieved in the previous steps.  We \textbf{included} all papers discussing the application of ML methods or improved ML pipeline procedures (e.g.\ data balancing), framing CSD as a classification task. We \textbf{excluded} papers only focusing on cross-version prediction (as this inherently implies that training and testing data has strong overlaps, and thus may cause overfitting). In case conference and extended journal variants exist of the same work, we only consider the extended (and more recent) journal variants.

\subsection{Review Results}
\label{sec:review}

Based on the process described above, we selected 45 papers: 25 papers for DP, four for PMT, and 16 for CSD, as presented in Table~\ref{tab:survey}. We observe the there is a large variability of strategies used to evaluate ML methods. The most common validation mechanism is n-fold cross-validation (18 papers\footnote{Two papers use a stratified variant of the n-fold cross validation}), followed by cross-project (12 papers), out-of-sample bootstrap (5 papers), and cross-version (3 papers) validation. Two articles did not specify what validation procedure was used. Four papers (i.e., \cite{amasaki2020cross, zhang2018predictive, khomh11bayesian, maiga12smurf}) apply more than one validation mechanism to assess the investigated ML methods. 

Only 22 out of 45 selected papers validated ML methods by performing multiple data-splitting repetitions in training and test sets. This, even though a prior DP study did indicate that the random nature of sampling used for the validation can introduce critical biases~\cite{Tantithamthavorn2017}, and different re-samplings can indeed lead to different results. For the remaining 23 papers, the authors did not mention any sampling repetitions, or acknowledge the absence of this in the threats to validity.

The results for the random seed are even more alarming. Only four out of 45 papers (8\%) rerun ML models with different random seeds. The only few papers that did so are related to DP, while no paper for the other two tasks rerun ML methods with multiple random seeds. For example, the heavily-used random forest, which even has been recommended as \textit{ most effective classifier in terms of performance}~\cite{mao2019extensive}, will by definition be sensitive to different random seeds.
While systematic and statistically sound methodologies to assessed randomized algorithms are well-established in other SE sub-domains~\cite{arcuri2014hitchhiker}, statistical tests on observed performance outcomes in predictive SE are only very rarely being mentioned.

The most widely used ML library is Weka~\cite{weka} (20 papers), followed by R~\cite{RManual} (8 papers), Scikit-learn~\cite{scikit-learn} (5 papers), and MATLAB~\cite{matlab} (3 papers). Keras and Tensorflow are the leading libraries for approaches based on Deep Learning. Finally, six papers did not report the actual ML library or framework used in the empirical evaluation. This raises some concerns about future replicability.

As for hyperparameter tuning, reporting the parameter values is not a common practice in predictive SE. Out of 45 papers, 29 (64\%) did not report the parameter values. 11 papers mention to have used
``default parameters''. 9 papers performed parameter tuning, but often did not reported the final (tuned) parameter values. Only one paper reported the tuned values~\cite{mao2019extensive}, and other two papers did so but partially~\cite{arcellifontana2016smell, barbez2020antipatterns}. 8 papers give indications of concrete parameter values being used; 8 other papers explicitly mention that a selection of parameter values are reported, or that reports are only provided for selected ML methods.

\begin{table*}[!htb]
    \centering
    \resizebox{\textwidth}{!}{
    \begin{tabular}{|lcccllc|cp{60pt}}
    \hline
    \textbf{Defect Prediction} & \textbf{Evaluation Type} & \textbf{\# Re-samplings} & \textbf{\# Random Seeds} & \textbf{Library} & \textbf{Parameter values} & \textbf{Notes} \\
    \hline
    Kondo et al., EMSE, 2019~\cite{kondo2019impact} & Out-of-sample bootstrap & 100 & Not Reported & R + Weka & Partially Tuned, Not Reported & Tuned only supervised methods \\
    Fan et al., TSE, 2019~\cite{fan2019impact} & Out-of-sample bootstrap & 1000 & Not Reported & R & Not Reported & ---\\
    Chen et al., IEEE Access, 2019~\cite{chen2019deepcpdp} & Cross-project (29\%-71\% split) & Not Reported & 10 & Keras & Partially Reported & Keras with default parameters\\
    Zhou et al., IST, 2019~\cite{zhou2019improving} & 2-fold cross-validation & 100 & Not Reported & Scikit-learn & Partially Reported & Baselines with default parameters\\
    Liu et al., IST, 2019~\cite{liu2019two} & Cross-project & 10 & 10 & LIBLINEAR + Weka & Not reported & SMOreg with default parameters \\
    Xu et al, IST, 2019~\cite{xu2019software} & Stratified sampling & 30 & Not Reported & MATLAB & Reported & ---\\
    Huang et al., EMSE, 2019~\cite{huang2019revisiting} & 10-fold cross-validation & 10 & Not Reported & Weka + R & Not Reported & Default parameters \\ 
    Ni et al., JSS, 2019~\cite{ni2019empirical} & Stratified sampling & 10 & Not Reported & Weka & Reported & ---\\
    Shippey et al., IST, 2019~\cite{SHIPPEY2019142} & Stratified cross-validation & 100 & Not Reported & Weka & Not Reported & Default Parameters \\
    Hoang et al, MSR, 2019~\cite{Hoang2019} & Cross-Validation & Not Reported & Not Reported & & Reported & ---\\ 
    Yu et al., IEEE Access, 2019~\cite{yu2019} & Cross-project & Not Reported & Not Reported & Weka & Reported & ---\\
    Bennin et al., EMSE, 2019~\cite{bennin2019relative} & 10-fold cross-validation & 10 & Not Reported & MATLAB + R & Not Reported & ---\\
    Zhang et al., IEEE Access, 2019~\cite{Zhang:access2019} & Cross-version & Not Reported & Not Reported & Weka & Not Reported & Default Parameters\\
    Gong et al., IEEE TR, 2020~\cite{Gong2020} & Cross-project & Not Reported & 20 & unknown & Not Reported & ---\\
    Cabral et al., ICSE, 2019~\cite{Cabral2019} & Cross-version & 30 & Not Reported & unknown & Tuned, Not Reported & ---\\
    Mori et al., EMSE, 2019~\cite{mori2019balancing} & 5-fold cross-validation & 5 & Not Reported & Weka + R & Reported & Reported seeds for data splitting\\
    Chen et al., IST, 2019~\cite{CHEN2019161} & 5-fold cross validation & 25 &  Not Reported & scikit-learn & Tuned, Not Reported & ---\\
    Jiarpakdee et al., TSE, 2019~\cite{Jiarpakdee2019} & Out-of-sample bootstrap & 100 & Not Reported & R & Partially Reported & Default values\\
    Jiarpakdee et al., TSE, 2019~\cite{Jiarpakdee} & Out-of-sample bootstrap & 100 & Not Reported & R & Partially Reported & Default values \\
    Chen et al., STVR, 2020~\cite{chen2020different} & Cross-project & Not Reported & Not Reported & Weka & Not Repored & Default values and ack. the diff. across libraries\\
    Chen et al., TSE, 2020~\cite{Chen2020empirical} & Cross-project & 20 & Not Reported & MATLAB & Not Reported & Friedman test + Nemenyi Test \\
    Amasaki, EMSE, 2020~\cite{amasaki2020cross} & Cross-project and version & Not Reported & Not Reported & CrossPare~\cite{herbold2015crosspare} & Not Reported & ---\\
    Afric et al., JSS, 2020~\cite{afric2020repd} & Random Splits & 30 & 30 & Scikit-learn & Not Reported & Default Values \\
    Yuan et al., IEEE Access, 2020~\cite{Yuan2020altra} & 2-fold cross-validation & 10 & Not Reported & Scikit-learn + CrossPare~\cite{herbold2015crosspare} & Not Reported & Default Values \\
    Alqadi \& Maletic, SANER, 2020~\cite{Alqadi2020slice} & Out-of-sample bootstrap & 1000 & Not Reported & R & Not Reported & --- \\
    \hline
    \hline
    \textbf{Predictive Mutation Testing}& \textbf{Evaluation Type} & \textbf{\# Re-samplings} & \textbf{\# Random Seeds} & \textbf{Library} & \textbf{Parameter values} & \textbf{Notes} \\
    \hline
   Zhang et al., ISSTA, 2016 ~\cite{zhang2016predictive} & Cross-project & NA & Not Reported & Weka & Not Reported & ---\\
   Zhang et al., TSE, 2018 ~\cite{zhang2018predictive} & Cross-version, cross-project & NA & Not Reported & Weka & Not Reported & ---\\
   Mao et al., ICST, 2019~\cite{mao2019extensive} & Cross-project 5-Folds & Not Reported & Not Reported & Scikit-learn & Tuned, Reported & Multiple reps. only for feature selection \\
    Naeem et al., IEEE Access, 2019~\cite{naeem2019scalable} & Cross-project & NA & Not Reported & Keras & Tuned, Not Reported & --- \\
    \hline
    \hline
    \textbf{Code Smell Detection}& \textbf{Evaluation Type} & \textbf{\# Re-samplings} & \textbf{\# Random Seeds} & \textbf{Library} & \textbf{Parameter values} & \textbf{Notes} \\
    \hline
    Kreimer, ENTCS, 2005~\cite{kreimer2005designflaws} & Leave-one-out & NA & Not Reported & Weka & Not Reported & Small-scale prototype \\
    Vaucher et al., WCRE, 2009~\cite{vaucher09godclass} & 10-fold cross-validation & Not Reported & Not Reported & Weka & Not Reported & Focus on rule extraction\\
    Hassaine et al., QUATIC, 2010~\cite{hassaine10ids} & 3-fold cross-validation  & Not Reported & Not Reported & Not Reported & Not Reported & No Variability Indicators\\
    Maneerat \& Muenchaisri, JCSSE, 2011~\cite{maneerat11badsmell} & 10-fold cross-validation & ``at least 10 times" & Not Reported & Weka 3.6 & Not Reported & Default parameters, no Variability Indicators\\
    Khomh et al., JSS, 2011~\cite{khomh11bayesian} & 3-fold cross-validation / cross-project & Not Reported & Not Reported & Weka & Not Reported & No Variability Indicators\\
    Maiga et al., WCRE, 2012~\cite{maiga12smurf} & 10-fold cross-validation / cross-project & Not Reported & Not Reported & Weka & Not Reported & No Variability Indicators\\
    Maiga et al., ASE, 2012~\cite{maiga12svm} & Not Reported & Not Reported & Not Reported & Weka & Not Reported & No Variability Indicators\\
    Amorim et al., ISSRE, 2015~\cite{amorim2015decisiontrees} & 10-fold cross-validation & Not Reported & Not Reported & Not Reported & Not Reported & No Variability Indicators\\
    Arcelli Fontana et al., EMSE, 2016~\cite{arcellifontana2016smell} & 10-fold cross-validation & 10 & Not Reported  & Weka & Tuned, Partially Reported & Standard deviation reported\\
    Arcelli Fontana \& Zanoni, KBS, 2017~\cite{arcellifontana2017smellseverity} & Stratified 10-fold cross-validation & 10 & Not Reported & Weka, KNIME & Taken From Previous Work & Standard deviation reported \\
    Fakhoury et al., SANER, 2018~\cite{fakhoury2018linguisticsmell} & Leave-one-out & NA & Not Reported & Weka, Tensorflow & Tuned, Reported & ---\\
    Di Nucci et al., SANER, 2018~\cite{dinucci2018smell} & Same as~\cite{arcellifontana2016smell} & Same as~\cite{arcellifontana2016smell} & Same as~\cite{arcellifontana2016smell} & Same as~\cite{arcellifontana2016smell} & Same as~\cite{arcellifontana2016smell} & No Variability Indicators\\
    Barbez et al., JSS, 2020~\cite{barbez2020antipatterns} & Cross-project & 1 & Not Reported & Tensorflow & Tuned, Partially Reported & --- \\
    Guggulothu \& Moiz~\cite{guggulothu2020multilabel}, SQJ, 2020 & 10-fold cross-validation & 10 & Not Reported & Not Reported & Not Reported & No Variability Indicators\\
    Pecorelli et al., JSS, 2020~\cite{pecorelli2020balancing} & 10-fold cross-validation & Not Reported & Not Reported  & Not Reported & Tuned, not reported & Classifier library not explicitly mentioned.\\
    Liu et al., TSE, 2020~\cite{liu2020deepsmell} & Cross-project & 1 & Not Reported & Keras & Not Reported & --- \\
    \hline
    \end{tabular}
    }
    \caption{Papers in the related literature that focus on defect prediction, predictive mutation testing, and code smell detection. 
    }
    \label{tab:survey}
\end{table*}

\section{Empirical Study}

As evidenced in our literature review, potential performance variability in predictive SE models is not frequently acknowledged.
However, beyond~\cite{Tantithamthavorn2017}, focusing on DP, no clear insight exists yet into the extent to which these factors have impact across predictive SE tasks. For example, the commonly used $n$-fold cross-validation scheme internally averages over $n$ train-test cycles, which is assumed to increase robustness and generalizability. Hence, it is unclear to what extent it practically matters that the initial partitioning into $n$ folds is randomized. Therefore, we conduct an empirical study, focusing on three research questions:
\begin{itemize}
    \item{\textbf{RQ1}: \textit{To what extent does the performance of cross-validated ML models vary when considering random fold repartitionings of the same dataset?}} 
    \item{\textbf{RQ2}: \textit{To what extent does the performance of randomized ML models vary when using different random seeds?}}
    \item{\textbf{RQ3}: \textit{To what extent are implementations of the same conceptual ML model in different libraries consistent?}}
    
\end{itemize}

\subsection{Datasets}
For our three predictive SE tasks of interest, we consider commonly used, openly available datasets\footnote{All data resources and repositories come from external, independent research. The exact input data files as used in our analyses are released in anonymized form as part of our supplementary material, available at \url{https://figshare.com/s/998dd6c07173f78bd64c}}.

\subsubsection{Defect Prediction (DP)} We use data from two repositories\footnote{From the combined repository at https://github.com/klainfo/DefectData.git.
}: PROMISE~\cite{boetticher2007promise}, and the repository built by D'Ambros et al.~\cite{d2010extensive}, which have been widely used in prior studies in defect prediction (e.g.,~\cite{kondo2019impact, amasaki2020cross, Gong2020, d2010extensive}). From the PROMISE repository~\cite{boetticher2007promise}, we considered six Java projects that are among the largest in terms of data points, that consider open-source projects (where applicable, according to latest versions), and that have representative data characteristics (i.e.\ defects not being the majority class): \texttt{Apache Ant} v1.7, \texttt{Camel} v1.6, \texttt{Ivy} v2.0, 
\texttt{jEdit} v4.3, \texttt{Lucene} v2.4, and \texttt{Apache Tomcat}.
From the repository built by D'Ambros et al.~\cite{d2010extensive}, we considered five available projects: \texttt{Eclipse JDT Core} 3.4, \texttt{Lucene} v2.4.0, \texttt{Mylyn} v3.1, \texttt{Eclipse PDE UI} v3.4.1, \texttt{Equinox} v3.4. The two repositories differ on the features (metrics) being used; further details are available in the original papers~\cite{menzies2006data, d2010extensive}.
The dataset sizes range from 341 (\texttt{Lucene} from PROMISE) to 2196 (\texttt{Mylyn} from D'Ambros et al.) data points (average dataset size: 877).

\subsubsection{Predictive Mutation Testing (PMT)} We use the data of the latest versions of the 9 base Git projects used in~\cite{zhang2018predictive}\footnote{\url{https://tinyurl.com/y6e4yyrr}}: \texttt{Java APNS}, \texttt{AssertJ}, \texttt{Joda-Time}, \texttt{Linear Algebra for Java}, \texttt{Apache Commons Lang}, \texttt{Message Pack for Java}, \texttt{UAAA}, \texttt{Vraptor}, and \texttt{Wire Mobile Protocol Buffers}. For feature details, we refer to the original paper~\cite{zhang2018predictive}. Dataset sizes range from 115 (\texttt{Java APNS}) to 32,930 (\texttt{Joda-Time}) data points (average dataset size: 10,278).

\subsubsection{Code Smell Detection (CSD)} We used the six datasets used in~\cite{arcellifontana2016smell}\footnote{\url{https://essere.disco.unimib.it/machine-learning-for-code-smell-detection/}}. Each dataset corresponds to one specific type of code smell (\textit{god class}, \textit{data class}, \textit{feature envy}, \textit{long method}, \textit{long parameter list}, and \textit{switch statements}). For feature details, we refer to the original paper~\cite{arcellifontana2016smell}. From each dataset, we remove columns with string values that are irrelevant to predictions (e.g., library/method names). Furthermore, we remove rows which contain missing values (indicated by a `?' in the original files), as the question of how to treat these missing values is not trivially defined.
Compared to the previous two tasks, the datasets for this task are smaller and within a closer range; dataset sizes range from 360 (\textit{long parameter list}) to 393 (\textit{data class}) data points.

\subsection{Study Design}
In our experiments, we frame our Predictive SE tasks of interest as binary classification tasks, and consider them under a 10-fold cross-validation setup (the most popular evaluation setup from Section~\ref{sec:review}), which will be rerun 100 times. To answer RQ1, for each dataset, 100 random 10-fold repartitionings are made; to answer RQ2, for each explicitly randomized ML model, 100 different, explicit random seeds are used.
As for the choice of ML models, we consider models that conceptually occur in three commonly used ML libraries (RQ3): \texttt{Weka}\footnote{\url{https://www.cs.waikato.ac.nz/ml/weka/}}, \texttt{scikit-learn} (sklearn) in Python, and the classifiers governed by the \texttt{mlr3}\footnote{\url{https://cran.r-project.org}} framework~\cite{mlr3} governing well-established ML packages in R. More specifically, we will consider Weka v3.8.4 in Java 11, sklearn v0.22.1 in Python 3.6, and mlr3 v0.5.0 in R v3.6.2.

Following this, for RQ1, we consider eight common supervised ML models: (1) Naive Bayes classifier (NB), (2) Linear Discriminant Analysis (LDA), (3) $k$-Nearest Neighbor classification (kNN), (4) Logistic Regression (LR), (5) Random Forest (RF), (6) Support Vector Machine (SVM), (7) Decision Tree (DT) and (8) Quadratic Discriminant Analysis (QDA). For RQ2, we investigated for which subset of these models explicit random seeds could be set in the libraries. Both in Weka and sklearn, these were the RF, SVM and DT models\footnote{In R, random state is implicit, and needs to be set as a global variable before calling the training procedure.}, and we therefore will focus on these three models in RQ3. 

 
%
%
\begin{table*}[!htb]
    \centering
    \tiny
    \begin{tabular}{|lp{60pt}p{350pt}l|}
    \hline
        \textbf{Library} & \textbf{Classifier Name} & \textbf{Default Parameters} & \textbf{References}  \\
    \hline
        Weka & \texttt{NaiveBayes} & \texttt{batchSize}=100, \texttt{numDecimalPlaces}=2, \texttt{useKernelEstimator}=false & \cite{witten2002data}\\
        \hline
        sklearn & \texttt{GaussianNB} & \texttt{priors}$^*$, \texttt{var\_smoothing}=1e-09 & --- \\
        \hline
        R & \texttt{naiveBayes} & \texttt{threshold}=0.001, \texttt{eps}=0 & \cite{e1071}\\
        \hline
    \hline
        Weka & \texttt{LDA} & \texttt{R}=1e-6 & \cite{witten2002data}\\
        \hline
        sklearn & \texttt{LinearDiscriminant}-\texttt{Analysis} &
            \texttt{solver} = `svd', \texttt{shrinkage}$^*$, \texttt{priors}$^*$, \texttt{n\_components} = min(n\_classes-1, n\_features),\texttt{ store\_covariance}=False, \texttt{tol}=0.0001 &
        \cite{hastie_et_al_statistical_learning,duda_hart}
        \\
        \hline
        R & \texttt{lda} & \texttt{prior}=proportions, \texttt{tol}=0.0001, \texttt{method}=`moment', \texttt{CV}=F & 
        \celllinebreak{
        \cite{MASS, ripley96}
        }\\
        \hline
    \hline
        Weka & \texttt{IBk} & \texttt{batchSize}=100, \texttt{numDecimalPlaces}=2, -k 1, -W 0, -A  `weka.core.neighbour-search.LinearNNSearch', -A \ `weka.core.EuclideanDistance', -R first-last & \celllinebreak{
        \cite{witten2002data, Aha1991}
        }\\
        \hline
        sklearn & \texttt{KNeighborsClassifier} & \celllinebreak{
            \texttt{n\_neighbors}=5, \texttt{weights}=`uniform', \texttt{algorithm}=`auto', \texttt{leaf\_size}=30, \texttt{p}=2,\\
            \texttt{metric}=`minkowski', \texttt{metric\_params}$^*$, \texttt{n\_jobs}$^*$
        } &
        --- \\
        \hline
        R & \texttt{kknn} & \texttt{k}=7, \texttt{distance}=2, \texttt{kernel}=`optimal', \texttt{ykernel}$^*$, \texttt{scale}=TRUE & \cite{kknn, hechenbichler2004kknn}\\
        \hline
    \hline
        Weka & \texttt{SimpleLogistic} & \texttt{batchSize}=100, \texttt{numDecimalPlaces}=2, -I 0 -M 500 -H 50 -W 0.0 & \cite{witten2002data, Landwehr2005, Sumner2005} \\
        \hline
        sklearn & \texttt{Logistic}-\texttt{Regression} &
            \texttt{penalty}=`l2', \texttt{dual}=False, \texttt{tol}=0.0001, \texttt{C}=1.0, \texttt{fit\_intercept}=True,
            \texttt{intercept\_scaling}=1, \texttt{class\_weight}$^*$, \texttt{random\_state}$^*$, \texttt{solver}=`lbfgs', \texttt{max\_iter}=100, \texttt{multi\_class} =`auto', \texttt{verbose}=0, 
            \texttt{warm\_start}=False, \texttt{n\_jobs}$^*$, \texttt{l1\_ratio}$^*$
            &
        \cite{wiki:bfgs, blog:bfgs} \\
        \hline
        R & \texttt{glm} & \texttt{weights}=rep.int(1, nobs),
        \texttt{start}$^*$, \texttt{etastart}$^*$, \texttt{mustart}$^*$,
         \texttt{offset}=rep.int(0, nobs), \texttt{family}=gaussian(), \texttt{intercept}=T, \texttt{singular.ok}=T & \cite{RManual}\\
        \hline
    \hline
        Weka & \texttt{RandomForest} & \texttt{batchSize}=100, \texttt{numDecimalPlaces}=2, -P 100, -I 100, -num-slots 1 -K 0, -M 1.0, -V 0.001, -S 1 & \cite{witten2002data, rf}\\
        \hline
        sklearn & \texttt{RandomForestClassifier} & 
            \texttt{n\_estimators} = 100, \texttt{criterion} = `gini', \texttt{max\_depth}$^*$,
            \texttt{min\_samples\_split} = 2, \texttt{min\_samples\_leaf} = 1,
            \texttt{min\_weight\_fraction\_leaf} = 0.0, \texttt{max\_features} = sqrt(n\_features),
            \texttt{max\_leaf\_nodes}$^*$, \texttt{min\_impurity\_decrease} = 0.0,
            \texttt{min\_impurity\_split}$^*$,
            \texttt{bootstrap} = True, \texttt{oob\_score} = False,
            \texttt{n\_jobs}$^*$, \texttt{random\_state}$^*$, \texttt{verbose} = 0, \texttt{warm\_start} = False, \texttt{class\_weight}$^*$, \texttt{ccp\_alpha} = 0.0, \texttt{max\_samples}$^*$
        &
        \cite{rf} \\
        \hline
R & \texttt{ranger} & \texttt{num.trees}=500,
\texttt{mtry}$^*$,
\texttt{importance}="none",
\texttt{write.forest} =T,
\texttt{probability} = F,
\texttt{min.node.size}$^*$,
\texttt{max.depth}$^*$, \texttt{replace} = T,
\texttt{sample.fraction} = ifelse(replace, 1, 0.632), 
\texttt{case.weights}$^*$,
\texttt{class.weights}$^*$,
\texttt{splitrule}$^*$,
\texttt{num.random.splits} = 1,
\texttt{alpha} = 0.5,
\texttt{minprop} = 0.1,
\texttt{split.select.weights}$^*$, 
\texttt{always.split.variables}$^*$, 
\texttt{respect.unordered.factors}$^*$, \texttt{scale.permutation.importance} = F,
\texttt{local.importance} = F, 
\texttt{regularization.factor} = 1, 
\texttt{regularization.usedepth} = F,
\texttt{keep.inbag} = F,
\texttt{inbag}$^*$,
\texttt{holdout} = F,
\texttt{quantreg} = F,
\texttt{oob.error} = T,
\texttt{num.threads}$^*$, 
\texttt{save.memory} = F, 
\texttt{verbose} = T,
\texttt{dependent.variable.name}$^*$, \texttt{status.variable.name}$^*$, 
\texttt{classification}$^*$,
& \cite{ranger, rf} \\
        \hline
    \hline
        Weka & \texttt{SMO} & \texttt{batchSize} = 100, \texttt{numDecimalPlaces}=2, -C 1.0, -L 0.001, -P 1.0e12, -N 0, -V -1, -K `weka.classifier.functions.supportVector.PolyKernel', -E 1.0, -C 250007, -calibrator `weka.classifiers.function.Logistic', -R 1.0E-8, -M -1 -num-decimal-places 4
        & \cite{witten2002data, Platt1998, Keerthi2001, Hastie1998}\\
        \hline
        sklearn & \texttt{SVC} &
            \texttt{C}=1.0, \texttt{kernel}=`rbf', \texttt{degree}=3, \texttt{gamma}=`scale', \texttt{coef0}=0.0,
            \texttt{shrinking}=True, \texttt{probability}=True, \texttt{tol}=0.001, \texttt{cache\_size}=200,
            \texttt{class\_weight}$^*$, \texttt{verbose}=False, \texttt{max\_iter}=-1, \texttt{decision\_function\_shape}=`ovr', \texttt{break\_ties}=False, \texttt{random\_state}$^*$
        &
        \cite{libsvm, platt99} \\
        \hline
        R & \texttt{svm} & \texttt{scale}=T, \texttt{type}$^*$, \texttt{kernel}="radial", \texttt{degree}=3, \texttt{gamma}=if (is.vector(x)) 1 else 1 / ncol(x), \texttt{coef0}=0, \texttt{cost}=1, \texttt{nu}=0.5, \texttt{class.weights}$^*$, \texttt{cachesize}=40, \texttt{tolerance}=0.001, \texttt{epsilon}=0.1,
\texttt{shrinking}=T, \texttt{cross}=0, \texttt{probability}=F, \texttt{fitted}=T
& \cite{e1071, libsvm}\\
        \hline
    \hline
        Weka & \texttt{J48} & \texttt{batchSize}=100, \texttt{numDecimalPlaces}=2, -C 0.25, -M 2 & \cite{witten2002data, Quinlan1993}\\
        \hline
        sklearn & \texttt{DecisionTreeClassifier} & 
            \texttt{criterion} = `gini', \texttt{splitter} = `best', \texttt{max\_depth}$^*$, \texttt{min\_samples\_split} = 2,
            \texttt{min\_samples\_leaf} = 1, \texttt{min\_weight\_fraction\_leaf} = 0.0, \texttt{max\_features}$^*$,
            \texttt{random\_state}$^*$, \texttt{max\_leaf\_nodes}$^*$, \texttt{min\_impurity\_decrease} = 0.0, \texttt{min\_impurity\_split}$^*$, \texttt{class\_weight}$^*$, \texttt{presort}=`deprecated', \texttt{ccp\_alpha}=0.0
        &
        \cite{hastie_et_al_statistical_learning, cart} \\
        \hline
        R & \texttt{rpart} & \texttt{method}=`class', \texttt{minsplit}=20, \texttt{minbucket}=round(minsplit/3), \texttt{cp}=0.01, \texttt{maxcompete}=4, \texttt{maxsurrogate}=5, \texttt{usesurrogate}=2, \texttt{xval}=10, \texttt{surrogatestyle}=0, \texttt{maxdepth}=30 \texttt{parms}=`gini' & \cite{rpart,cart}\\
        \hline
    \hline
        weka & \texttt{QDA} & \texttt{R}=1e-6  & \cite{witten2002data}\\
        \hline
        sklearn & \texttt{QuadraticDiscriminant}-\texttt{Analysis} &
        \celllinebreak{
            \texttt{priors}$^*$, \texttt{reg\_param}=0.0, \texttt{ store\_covariance}=False, \texttt{tol}=0.0001} &
        \celllinebreak{
        \cite{hastie_et_al_statistical_learning, duda_hart}
        } 
        \\
        \hline
        R & \texttt{qda} & \texttt{prior}=proportions, \texttt{tol}=0.0001, \texttt{method}=`moment', \texttt{CV}=F & \cite{MASS, ripley96}\\
    \hline  
    \multicolumn{4}{c}{Parameters marked with {*} are set equal to NULL (in R) or None (in Python)}
    \end{tabular}
    \caption{Specifications and used (default) settings of the considered classifiers for the three libraries of interest. In case library documentation explicitly References work related to implementation and default parameter configurations, we list those here.}
    \label{tab:classifier_defaults}
\end{table*}

To highlight potential differences between libraries, we fix and store data partitionings and random seed choices for the 100 runs once\footnote{Exact configurations are shared as part of the supplementary material at \url{https://figshare.com/s/998dd6c07173f78bd64c}}, and reuse these for all libraries. We will run the classifiers with their default parameters, unless our experiments explicitly demand a specific parameter setting\footnote{For the RQ2 experiments, we override default random seed parameters explicitly; furthermore, in order to be able to evaluate AUC scores, we explicitly set SVM classifiers in sklearn to output probability estimates.}.
In doing this, we purposefully mimic a scenario of naively running the ML models in their out-of-the-box form; as our review in Section~\ref{sec:review} showed, this is not uncommon in literature. Details of exact classifier names and default parameter values in all libraries of interest are given in Table~\ref{tab:classifier_defaults}.

To minimize the researcher degrees of freedom and maximize reproducibility, we also minimize preprocessing of the data in the datasets,
and purposefully do not apply any rebalancing or normalization steps, nor any hyperparameter tuning.
We emphasize that in none of our approaches, we seek to optimize for performance explicitly; instead, we seek to quantify natural performance variability.

For RQ1, we will run 26 (datasets) $\times$ 8 (ML models) = 208 experiments per ML library, performing 100 10-fold cross-validations for each experiment. To answer RQ3, we repeat this procedure for Weka, sklearn, and R. As we reuse the same pre-generated 100 10-fold cross-validation partitionings, the $i$-th experimental run for any ML library will have been based on the exact same randomized partitioning for the 10-fold cross-validation. For the 208 $\times$ 3 (libraries) experiments, we calculate the performance median and range, according to 3 metrics, resulting in 1872 median-range pairs.

For RQ2, for each dataset, we take the first 10-fold partitioning considered as part of RQ1, and then run the cross-validation 100 times, now considering different, explicitly set random seeds in each run. In total, for this RQ, we run
26 $\times$ 3 = 78 experiments, each encompassing 100 runs, wherein the $i$-th experimental run, we use $i$ as an explicit random seed to be passed to the three randomized ML models of interest\footnote{In Weka, this also meant that we had to set the \texttt{ReducedError}-\texttt{Pruning} parameter to true, to allow J48 using different seeds.}. 
Repeating this for all three ML libraries of interest, we obtain 78 $\times$ 3 $\times$ 3 = 702 median-range pairs.

\subsection{Evaluation metrics}
For all three RQs, we consider three evaluation metrics: the commonly applied (1) F-measure and (2) Area Under the Curve (AUC), plus the (3) Matthews Correlation Coefficient~\cite{Shepperd2014bias}. F-measure requires to set a cut-off (usually set to 0.5) on the confidence levels produced by ML methods to decide whether a given data point should be marked as a true case (e.g., defective class in defect prediction) or not.  Instead, the AUC metric does not require a cut-off and, therefore, it is often preferred to other metrics~\cite{Kamei2016}. Finally, the Matthews Correlation Coefficient (unlike the F-measure and AUC) uses all four quadrants of the confusion matrix, and has been advocated as a more stable metric in the recent DP literature~\cite{Shepperd2014bias}. To compute these metrics, we make use of the implementations offered by the various ML libraries. 

For our analysis, we first consider the median performance (i.e., F-measure, AUC, and Matthews Correlation Coefficient) achieved by each classifier on each dataset and across 100 independent runs (100 data resamplings for RQ1 and 100 random seeds for RQ2). To analyze the variability in the results across different runs,
we compute the \textit{range}. This is a standard dispersion metric in statistics, reflecting the difference between the largest and smallest values within a given distribution. In our case, this concerns the maximum performance difference of a given classifier on a given dataset, observed over the 100 independent runs for the RQ of interest; in other words, the difference between best-case and worst-case performance in particularly `(un)lucky' runs in our experiments. Due to space considerations, we will present observed median results and the range values through swarm plots\footnote{Complete outcomes in tabular form are offered as supplementary material at \url{https://figshare.com/s/998dd6c07173f78bd64c}}: scatter plots in which points corresponding to the same value are plotted in close proximity to one another, rather than on top of each other. This way, both the local density and overall spread in a distribution can be seen. Per RQ, we will plot medians and ranges for each choice of the dataset, ML model, metric and ML library. Results for datasets within the same predictive SE task will be included in the same plot.




To answer RQ3 from a statistical point of view, we
will not use the Scott-Knott Effect Size Difference proposed in~\cite{Tantithamthavorn2017}; instead, following the commentary in~\cite{herbold2017comments}, we will use the Friedman test~\cite{friedman1940comparison} with the Nemenyi post-hoc test~\cite{demvsar2006statistical}. The Friedman test is the non-parametric equivalent to ANOVA with
unreplicated block design~\cite{friedman1940comparison}. Due to its non-parametric nature, it does not make any assumption about the type of distribution (i.e., the data does not need to follow a normal distribution), and allows us to assess the significance of differences across multiple test attempts (in our case, datasets) when using different treatments (in our case, ML methods and their implementations). We use the confidence level $\alpha=0.95$~\cite{friedman1940comparison}.

While the Friedman test reveals whether results
statistically differ among different treatments, it does not tell for which combination of treatments the significance holds. To further understand this aspect, we use the Nemenyi post-hoc test~\cite{demvsar2006statistical}. The Nemenyi test computes the average rank achieved by each treatment across multiple datasets. Treatments with the best ranking are preferable over the others. Two treatments (e.g., RF in Weka vs.\ RF in R) significantly different if their corresponding average ranks differ by at least the given critical distance (CD)~\cite{demvsar2006statistical}.
\begin{figure*}[!tb]
    \centering
    \includegraphics[width=\linewidth]{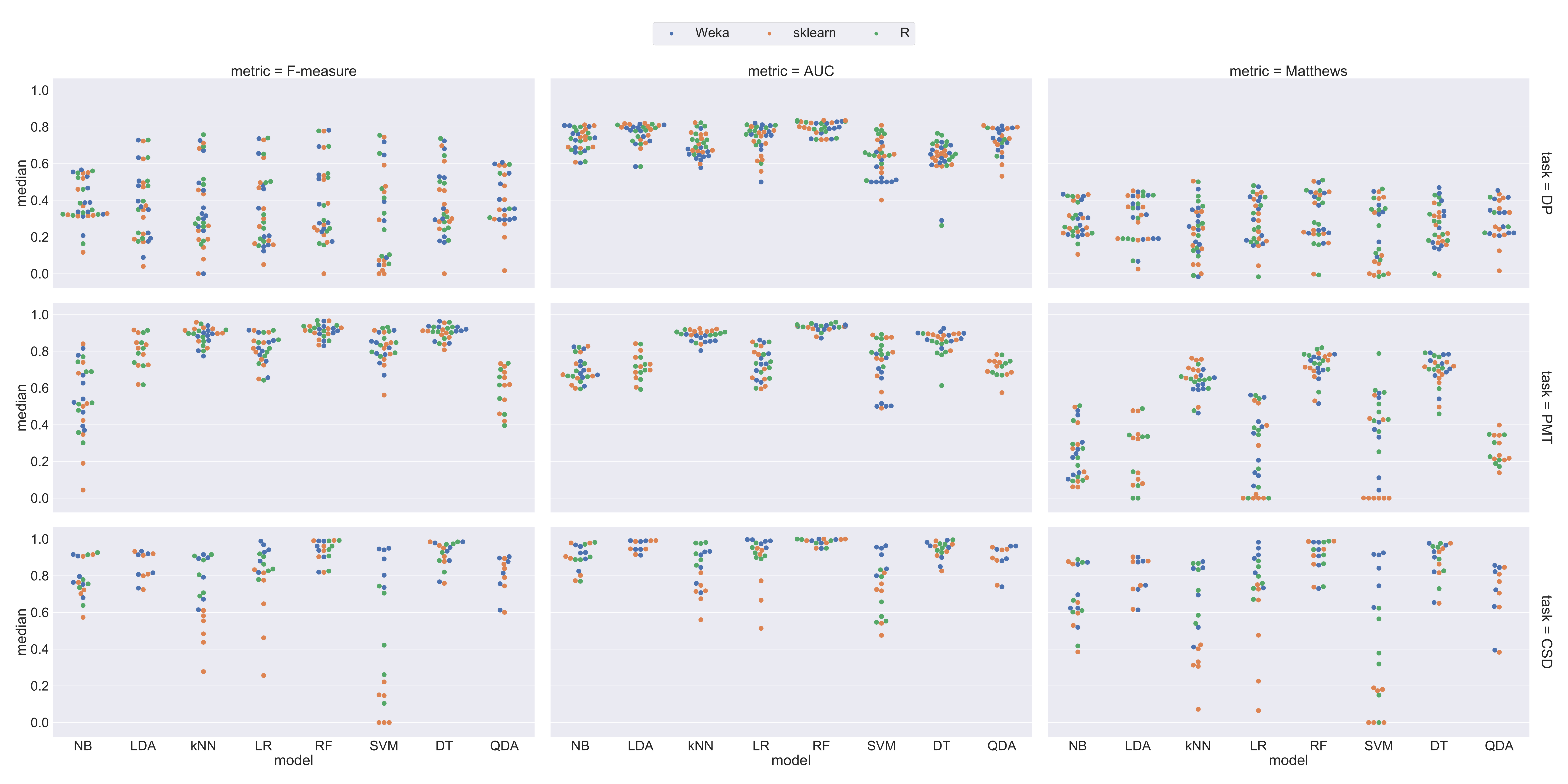}
    \vspace{-4mm}
    \caption{Swarm plots of median performance observed over 100 runs per dataset, using different random 10-fold cross-validation partitions.}
    \label{fig:RQ1_swarm_median}
\end{figure*}
\begin{figure}[!tb]
    \centering
    \includegraphics[width=\columnwidth]{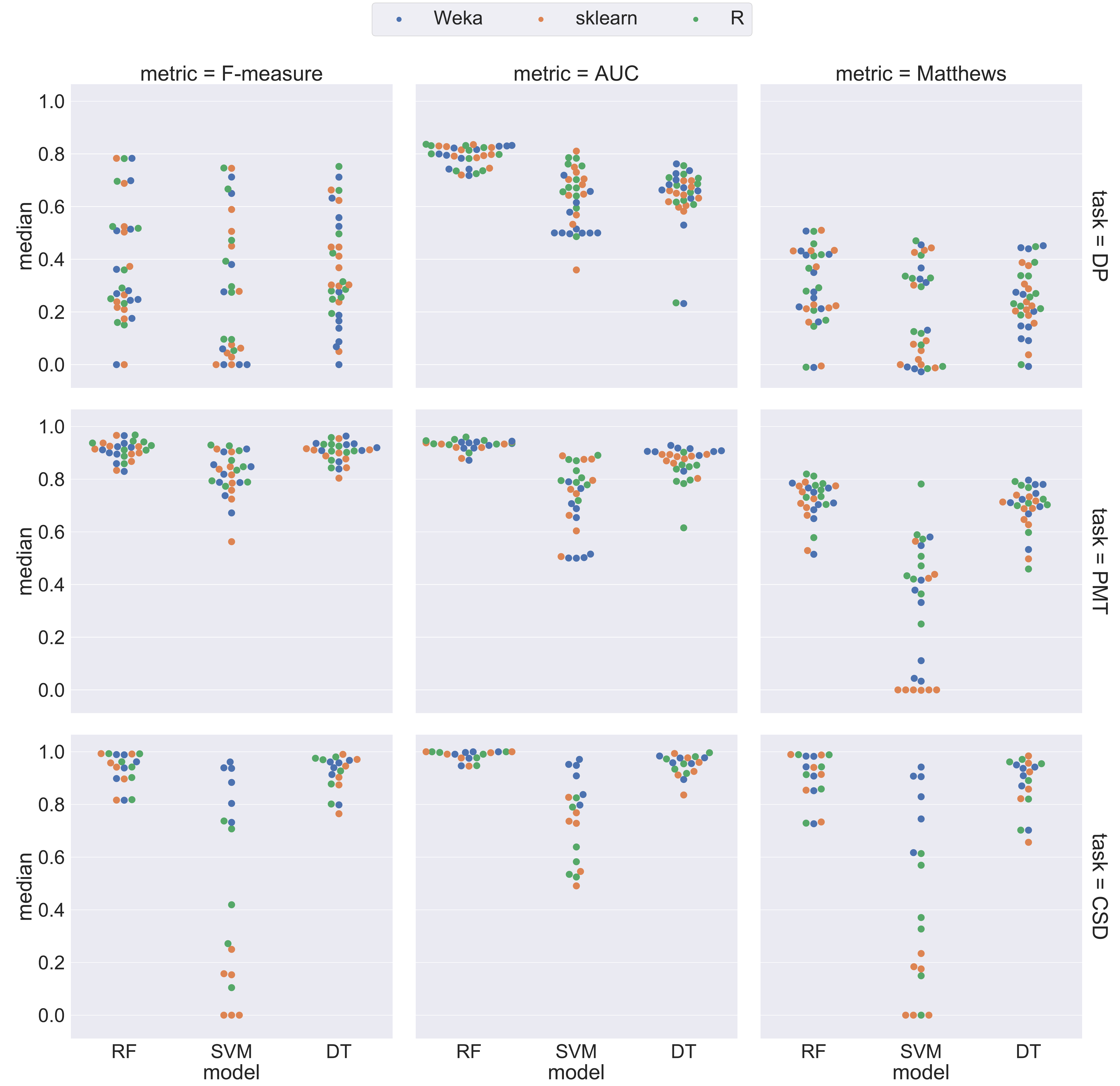}
    \vspace{-4mm}
    \caption{Swarm plots of median performance observed over 100 runs per dataset, using different random 10-fold cross-validation partitions.}
    \label{fig:RQ2_swarm_median}
\end{figure}

\begin{figure*}[!tb]
    \centering
    \includegraphics[width=\linewidth]{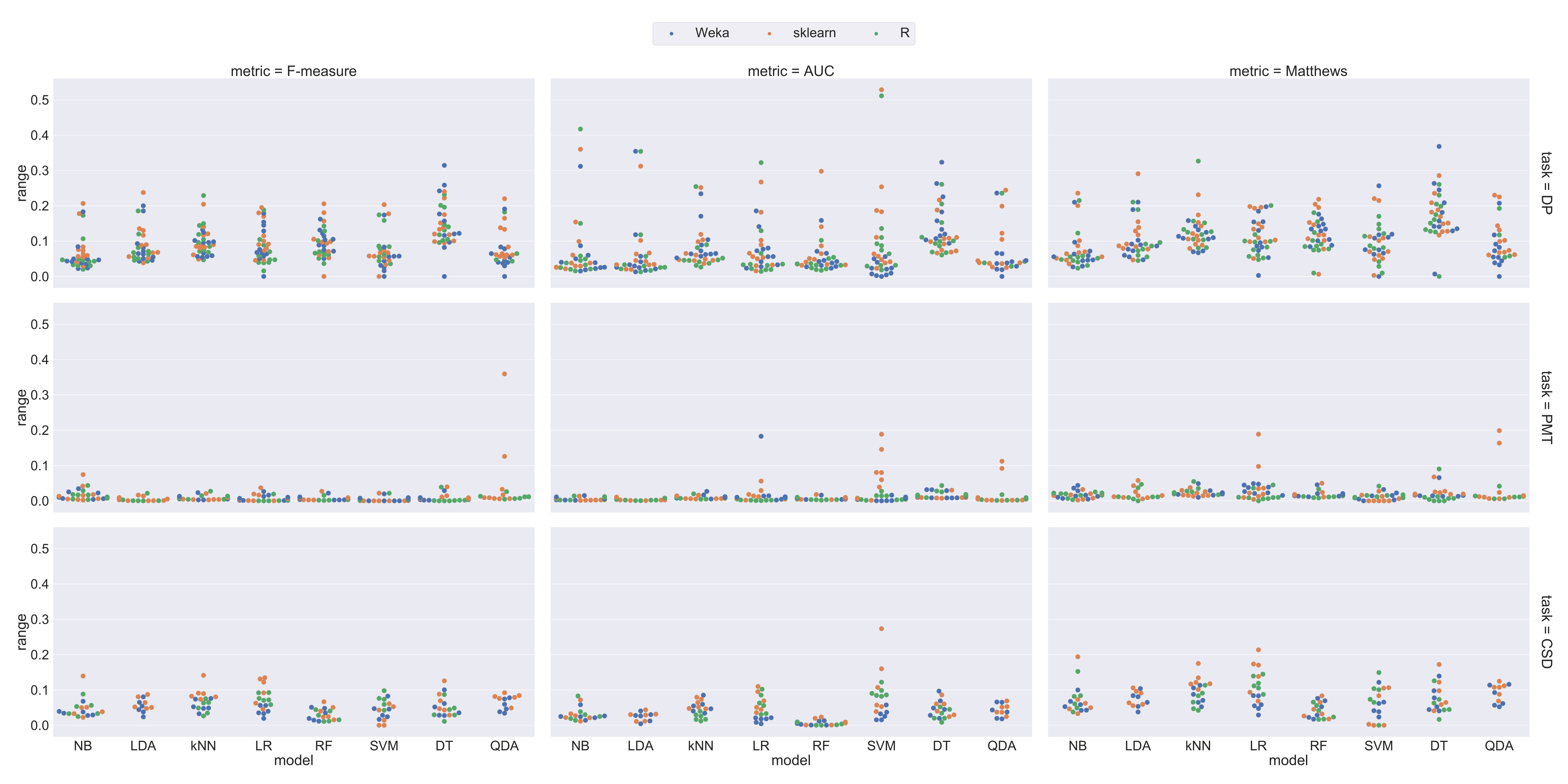}
    \caption{Swarm plots of performance ranges observed over 100 runs per dataset, using different random 10-fold cross-validation partitions.}
    \label{fig:RQ1_swarm_range}
\end{figure*}
\begin{figure}[!tb]
    \centering
    \includegraphics[width=\columnwidth]{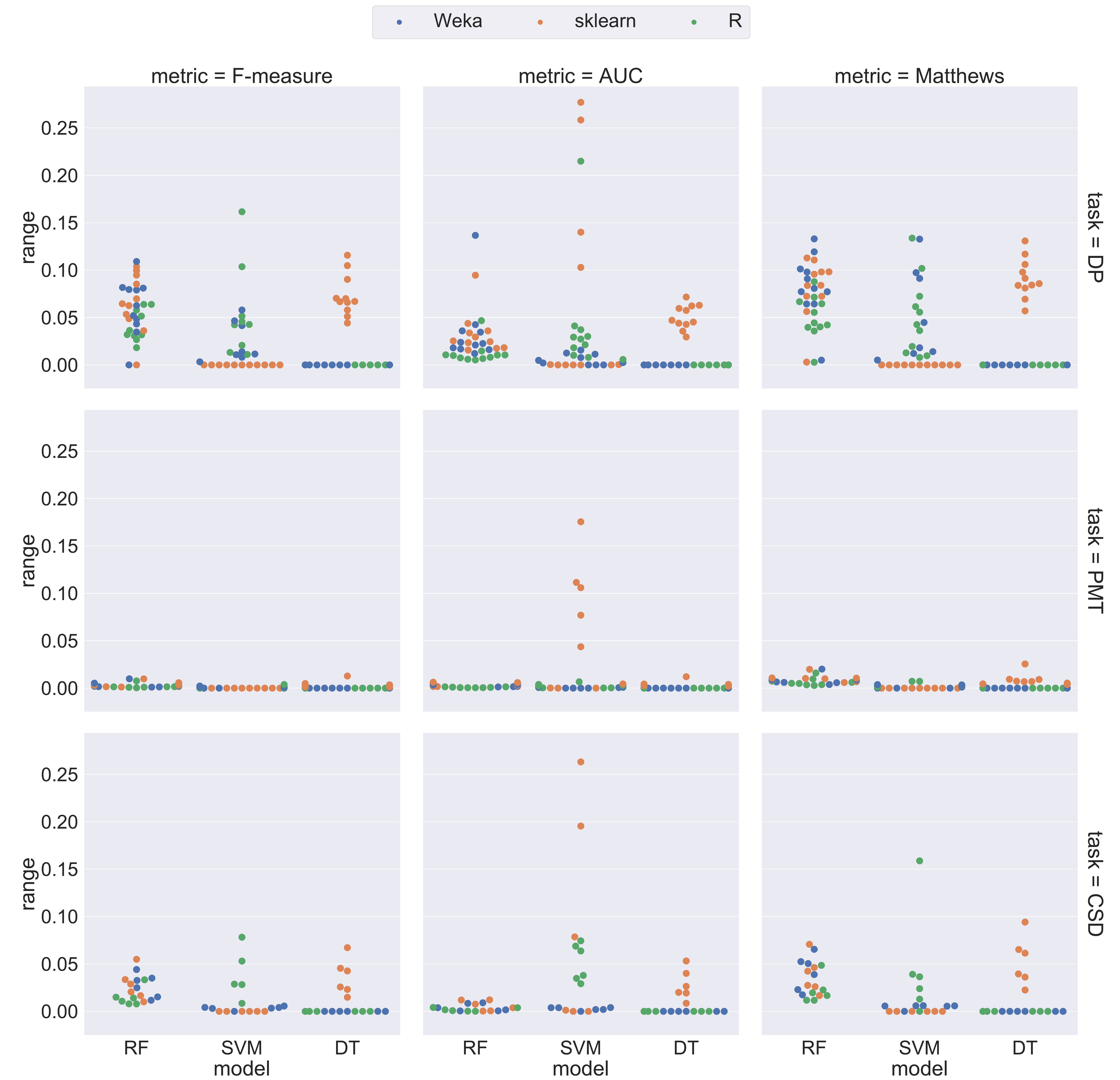}
    \caption{Swarm plots of performance ranges observed over 100 runs per dataset, using different random seeds for the ML training.}
    \label{fig:RQ2_swarm_range}
    \vspace{-5mm}
\end{figure}

\section{Empirical Results}\label{sec:results}


\subsection{Results}

\emph{Median performances.}
Median performance observations indicate what (absolute) performance outcomes are observed `in the middle' of the performance range, when considering multiple runs. Swarm plots for RQ1 and RQ2, both considering all three ML libraries (RQ3), are presented in Figure~\ref{fig:RQ1_swarm_median} and~\ref{fig:RQ2_swarm_median}, respectively. Comparing results for RQ1 and RQ2, we notice similar distributional patterns and ranges, and also note that performance for an ML model may greatly differ, depending on the considered dataset.

Considering Weka to be the most frequently used library in the field, even while the current models were not explicitly tuned and no domain-specific data pre-processing was performed, observed median performance numbers appear in alignment with performance reports in the existing literature (e.g.\ from the meta-analysis in~\cite{azeem19smellreview}). However, when not only looking at results from Weka, remarkably, different ML libraries appear to give inconsistent results. For example, observed performances for sklearn's kNN, LR and SVM classifiers appear to perform worse than the other two ML libraries for various datasets in the CSD task. Also, in some cases, for the same dataset and the same conceptual ML model, classifiers may crash or raise warnings in one library, but not the other. As one example, in R, no LDA and QDA outcomes can be obtained for any dataset in the CSD task, due to constant values being observed within groups for LDA, and rank deficiencies being observed for QDA. In sklearn, LDA completes without any warnings on the same inputs; indeed, collinearity warnings are given for QDA, but the experiments complete without crashing. In Weka, no crashes are observed, and no warnings are given.


\begin{figure*}[!tb]
    \centering
    \includegraphics[width=0.8\linewidth]{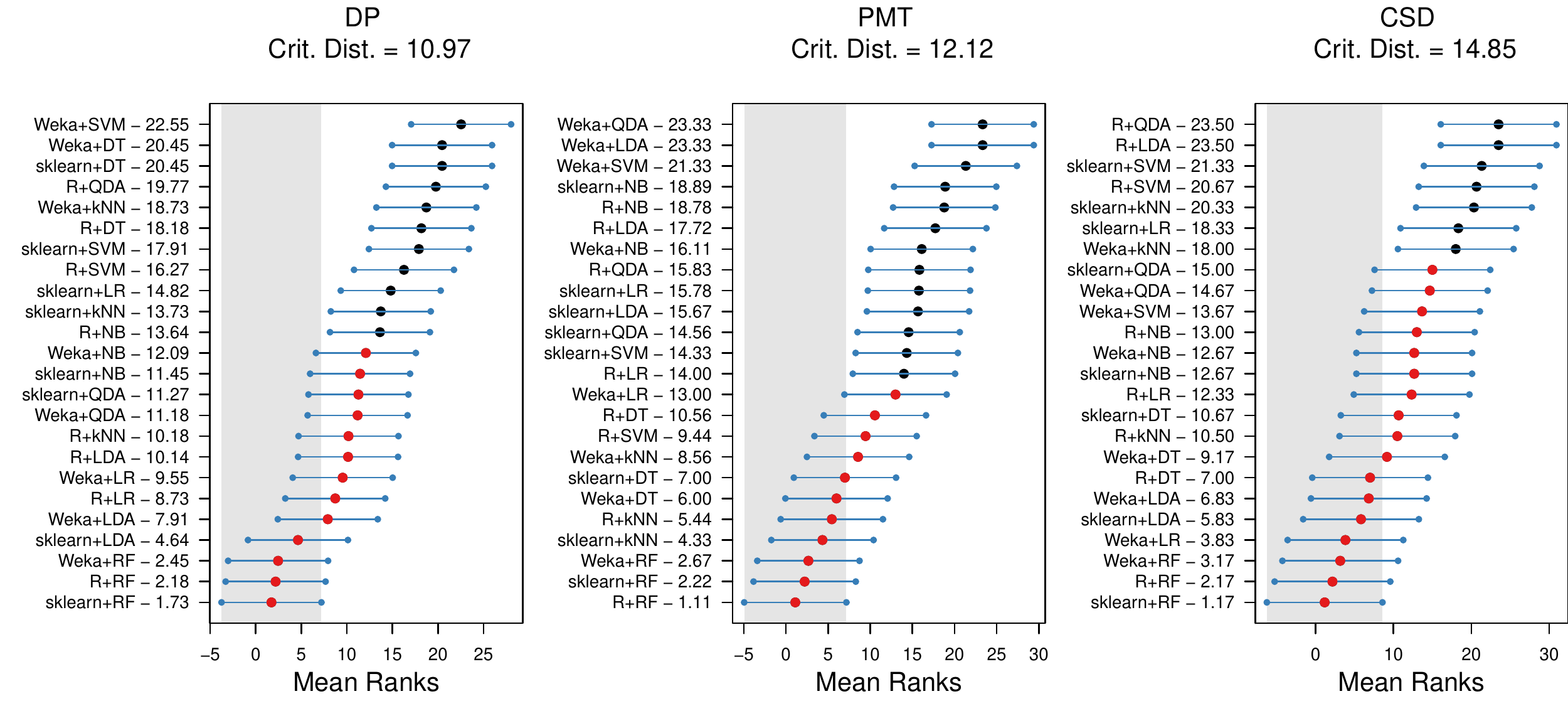}
    \caption{Results of the statistical comparison based on the Friedman test and the Nemenyi post-hoc test.}
    \label{fig:RQ3}
\end{figure*}

\emph{Performance ranges.}
To highlight differences between particularly good and bad runs, we calculate and analyze performance ranges. The performance ranges for RQ1 and RQ2 are shown in Figures~\ref{fig:RQ1_swarm_range} and~\ref{fig:RQ2_swarm_range}, respectively, again representing outcomes for all three ML libraries of interest (RQ3). 

We observe large ranges, confirming that the choice of data partition may cause considerable performance differences. Considering that the F-measure and AUC can range between $[0.0, 1.0]$ (while the Matthews Correlation Coefficient can range between $[-1.0, 1.0]$), observed performance ranges for RQ1 (Figure~\ref{fig:RQ1_swarm_range}) are particularly alarming for these two metrics: e.g., for the DP task, considering the AUC, in some cases, performance ranges higher than 0.4 are observed. The dataset for which the largest differences were observed was \texttt{jEdit} v4.3; upon closer inspection, we found ths dataset to be highly skewed (containing only 11 defective classes over a total of 492 classes in the project); as a consequence, considering 10-fold partitioning, there may have been folds without any true positives.

Comparing range results for RQ1 and RQ2 (Figure~\ref{fig:RQ2_swarm_range}), two interesting observations can be drawn for RQ2. First of all, in the absolute sense, performance ranges for runs based on different random seeds are smaller than those for runs based on different data repartitionings. Furthermore, as datasets grow larger in size, in line with statistical expectations, variability will decrease; this can e.g.\ be seen by comparing ranges observed for PMT (for which datasets were considerably larger in size) with those observed in the two other tasks.

Considering RQ3, it is striking that certain libraries yield (near-)deterministic results for different random seed choices, while others do not. For example, DT outcomes for Weka and R are very consistent, while for sklearn, they differ for different random seeds.

\emph{Multi-dataset Statistical Analysis.}
While we observed that median performance outcomes are in alignment with currently reported performance in literature, can any ML model be claimed to be superior over all other models? For this, as discussed before, we conduct a Friedman test and a Nemenyi post-hoc test, comparing model ranks, based on median AUC performance.
To immediately consider this question in the context of RQ3, we consider \emph{each combination of ML model and ML library} as a treatment. As different predictive SE tasks consider different problems, potentially with different data characteristics, we conduct separate statistical analyses per task.

Our results are shown in Figure~\ref{fig:RQ3}; several striking aspects stand out. First of all, while RF models consistently are high ranked for all three ML libraries, concerning the critical distance, they cannot be claimed as statistically superior to all other ML models. Next to this, while different ML libraries reimplement the same conceptual ML model, the default configurations of these models do not give equivalent results. This is evidenced by different libraries relating to the same model generally not grouping together in rankings; in some cases, e.g.\ for LDA and kNN, ranks may greatly differ.

\subsection{Discussion}
While in terms of median performance, results of our studied ML models appear to be in realistic ballparks, very large differences may occur for different randomized runs, especially when considering different data repartitionings. Furthermore, considering parallel implementations of ML models in various common libraries, we observe major inconsistencies in performance outcomes between libraries.

Several explanations can be found for these inconsistencies. First of all, while referring to the conceptually same ML model, underlying design choices may not actually implement the same model. For example, as for DT, the J48 decision tree implemented in Weka is a C4.5 tree~\cite{Quinlan1993}, while sklearn and R implement CART trees~\cite{cart}. Careful examination of Table~\ref{tab:classifier_defaults} also reveals major differences in default parameter settings for the different models. For example, for kNN, the default $k$ is 1 for Weka, 5 for sklearn and 7 for R; for SVM, the default kernel in Weka is polynomial, while it is a radial basis function in sklearn and R.

It should however be emphasized that no optimal default parameter setting exists. When looking at Figure~\ref{fig:RQ3}, considering alternative model implementations, no consistent order is found between our three ML libraries. That the RF models, even despite many different parameters in different libraries, still form the most consistent high-performing block in the rankings, can be attributed to them being an ensemble method, performing bootstrap aggregating on decision trees. As a consequence, while the underlying methods and parameters may differ, if many estimators are combined, results may converge and become more robust.

\section{Threats to Validity}

\textit{Internal Validity}. Several threats to internal validity may be identified. Firstly, despite our best effort applying procedural best practices~\cite{kitchenham2004, kitchenham2009systematic}, our systematic review may still miss relevant works. However, we sought to provide a representative and recent snapshot, rather than an exhaustive overview. Furthermore, patterns with regard to (the absence of) detail reporting were so strong, that even the addition of further works will be unlikely to alter these patterns.

Secondly, as we very explicitly used default ML model parameter values (without hyperparameter tuning), and did not apply any domain-specific data pre-processing, one can challenge to what extent our experimental runs considered realistic approaches to the predictive SE tasks of interest. However, as followed from our review in Section~\ref{sec:review}, the use of default values is not uncommon, and when hyperparameters are tuned, there is no obvious optimal parameter configuration to replicate, transfer and/or `plug in' trivially. Furthermore, our goal was not to solve each predictive SE task in the best way possible, but rather to give insights into what threats to reproducibility may already occur in minimal-intervention, out-of-the-box setups. Therefore, we feel the choice not to do domain-specific data pre-processing is justified, as doing this would lead to additional, non-trivially reported researcher degrees of freedom to be added to the procedure.


\textit{External Validity}. Questions may be raised regarding the generalizability of our findings. Our chosen datasets all come from well-established, openly available sources that are more broadly used in the field. In choosing the datasets, we also were careful to prioritize newer versions, larger datasets, and realistic projects. Still, it can be questioned to what extent our datasets representatively and comprehensively represent the predictive SE tasks of interest, also considering that all project-related datasets consider Java projects. However, our interest in randomization-induced performance variability and the comparison of parallel implementations of the same ML model actually is not dependent on the input data---in fact, we could have applied a similar procedure to any other binary classification task in any other field. However, with the importance of ML for predictive SE, we did choose to restrict to data from this field, and our setup can easily be expanded to further datasets and further tasks. 

\section{Practical Guidelines for Predictive SE}
\label{sec:guide}
%
We have shown that commonly underreported aspects of ML procedures can lead to large performance differences and unexpected outcome inconsistencies. From our findings, we can distill several practical guidelines that can help improve the reproducibility and statistical soundness of studies.    


\vspace*{2mm}
\hspace*{-5mm}
\begin{tikzpicture}
\node [mybox] (box){%
\centering
\begin{minipage}{.465\textwidth}
\textbf{Guideline 1}. Rerun ML models on many different resamplings of the same dataset. 
\end{minipage}
};
\end{tikzpicture}%

As we demonstrated, different random partitionings for 10-fold cross-validation may lead to strongly differing performance outcomes. Our review in Section~\ref{sec:review} showed that multiple fold repartitionings are rarely considered in cross-validation setups; if they are, they are rarely repeated more than 10 times. However, typically considered ML models and datasets in predictive SE, such as the ones we used in this paper, are not so large or complex that they demand extensive computational resources. Thus, it was inexpensive to perform 100 runs in this work, and we feel explicit justification is needed when researchers choose not to perform many reruns.

While our study focused on within-project cross-validation, repeated resampling is also possible for cross-project~\cite{Chen2020empirical}, and cross-version~\cite{Cabral2019} validation. Beyond taking full projects or versions as alternatives to folds in a cross-validation setup, combinations of versions and/or projects may be randomly drawn to increase an ML model's robustness. In such cases, we recommend using many repeated resamplings too.


\vspace*{2mm}
\hspace*{-5mm}
\begin{tikzpicture}
\node [mybox] (box){%
\centering
\begin{minipage}{.465\textwidth}
\textbf{Guideline 2}. Rerun ML models considering multiple, explicitly controlled random seeds. 
\end{minipage}
};
\end{tikzpicture}%
    
While observed performance ranges were smaller for RQ2 than for RQ1, different random seeds choices were still shown to cause non-negligible performance variability, which was also inconsistent for implementations of the same model in different libraries. While this is currently hardly acknowledged in the literature, we recommend to consciously rerun models on a given data split configuration with multiple, different, and explicitly controlled random seeds. For our study, performing 100 runs was practically feasible.

\vspace*{2mm}
\hspace*{-5mm}
\begin{tikzpicture}
\node [mybox] (box){%
\centering
\begin{minipage}{.465\textwidth}
\textbf{Guideline 3}. Be aware that seemingly non-randomized ML procedures may converge differently.
\end{minipage}
};
\end{tikzpicture}%


While explicit randomization only occurs for a few ML models, iterative numerical methods may be applied during the training phase. This can again lead to performance variability, depending on (1) the solver being used (randomized or not), (2) the number of iterations (if used as stopping criterion), and (3) the tolerance threshold. In our study, the default settings for NB, LDA, KNN, LR, and QDA (see Table~\ref{tab:classifier_defaults}) use deterministic solvers with very small tolerance values, such that consistent results were obtained for multiple reruns under the same data and parameter configuration. Researchers should explicitly be aware of this and check the configurations of the solvers.

In addition, similar considerations come into play when hyperparameters are tuned. While hyperparameter search procedures were sometimes reported in our review, explicit final outcomes rarely are. If paper assets would not allow such reports, we call for both the search procedure and final parameter choices to be explicitly included in replication packages.

    
\vspace*{2mm}
\hspace*{-5mm}
\begin{tikzpicture}
\node [mybox] (box){%
\centering
\begin{minipage}{.465\textwidth}
\textbf{Guideline 4}. Explicitly report parameter values, library/function names, and library versions.
\end{minipage}
};
\end{tikzpicture}%

As our results show, running default settings of the (conceptually) same ML model in different libraries yields very different results. Judging from these outcomes, generic statements like ``we trained an SVM on the openly available dataset X'' are irreproducible. While design choices and default settings differ between different ML libraries, as shown in Table~\ref{tab:classifier_defaults} and discussed in Section~\ref{sec:results}, we also emphasize that they may change across multiple versions of the same library. For example, for RF in sklearn, the default number of trees in the forest (the \texttt{n\_estimators} parameter) changed\footnote{\url{https://tinyurl.com/yyd4kqbu}} from 10 (versions $<$ 0.22) to 100 (version $\geq$ 0.22).



\vspace*{2mm}
\hspace*{-5mm}
\begin{tikzpicture}
\node [mybox] (box){%
\centering
\begin{minipage}{.465\textwidth}
\textbf{Guideline 5}. Consider which outcomes can truly be considered as statistically significant.
\end{minipage}
};
\end{tikzpicture}%

Only very few works in our survey made use of statistical tests to assess what differences can truly be deemed significant. Interest in statistically sound performance assessment is emerging in the SE domain~\cite{arcuri2014hitchhiker, Tantithamthavorn2017, herbold2017comments}, and for our current experiments, the Friedman test complemented with a post-hoc Nemenyi test (as recommended in~\cite{demvsar2006statistical} and~\cite{herbold2017comments}) was appropriate. For our work, a major realization following this procedure has been that \emph{while RF models consistently rank high in performance, from a statistical perspective, they cannot be considered fully superior to other ML models}. We recommend that the SE community more consciously take these types of tests into account when reporting and assessing performance outcomes.


\section{Conclusions and Future Work}
In this work, we provided a systematic review of common reporting practices regarding ML procedures in Predictive SE, indicating that randomization aspects are very rarely explicitly acknowledged in current work. We ran experiments with multiple randomized runs (in terms of fold partitioning and random seed choice) on representative datasets in three Predictive SE tasks, for eight common ML models implemented by three common ML libraries, and observed considerable performance differences. Outcomes from different library implementations of the conceptually same ML model were inconsistent, likely due to greatly differing default parameter choices. 

From our findings, it became clear that much more conscious awareness of these issues is needed, and that much more transparency in reporting is needed to make outcomes reproducible. To this end, we proposed several practical guidelines. Given the continued interest in ML for SE, and the increasing number of papers on ML observed in SE top venues, we hope these guidelines will help in improving best practices and strengthening the field.

\balance
\bibliographystyle{IEEEtran}
\bibliography{references}

\end{document}